\documentclass[conference]{IEEEtran}
\IEEEoverridecommandlockouts
\usepackage{cite}
\usepackage{amsmath,amssymb,amsfonts}
\usepackage{algorithmic}
\usepackage{graphicx}
\usepackage{textcomp}
\usepackage{xcolor}
\usepackage{tikz}
\usepackage{url}

\usepackage{ifthen}
\usepackage{soul}

\newboolean{highlight-rebuttal-change}
\setboolean{highlight-rebuttal-change}{false} 

\newcommand{\rbtadd}[1]{\ifthenelse{\boolean{highlight-rebuttal-change}}{\textcolor[HTML]{19BC09}{#1}}{#1}}
\newcommand{\rbtremove}[1]{\ifthenelse{\boolean{highlight-rebuttal-change}}{\textcolor[HTML]{FF4545}{\st{#1}}}{}}

\newcommand{\allimgwidth}{72mm}
\newcommand{\threesquareimgwidth}{68mm}

\def\BibTeX{{\rm B\kern-.05em{\sc i\kern-.025em b}\kern-.08em
    T\kern-.1667em\lower.7ex\hbox{E}\kern-.125emX}}
\newcommand{\swatch}[1]{\tikz[baseline=-0.6ex] \node[fill=#1,shape=rectangle,draw=darkgray,ultra thin,minimum width=3mm, rounded corners=2pt](){};}
\definecolor{bluish}{RGB}{128,128,255}

\begin{document}
\title{
    \rbtadd{Analysis and Compilation of }Normal Map Generation \rbtadd{Techniques }for Pixel Art\\
    \thanks{
        This work was partially supported by CAPES, CNPq and Fapemig\\
        978-1-6654-6156-6/22/\$31.00 ©2022 IEEE
    }
}

    
    

\author{\IEEEauthorblockN{Rodrigo D. Moreira}
    \IEEEauthorblockA{
        Departamento de Computação\\
        CEFET-MG\\
        Belo Horizonte, Brasil\\
        rodrigodmoreira.rdm@gmail.com
    }
    \and
    \IEEEauthorblockN{Flávio Coutinho}
    \IEEEauthorblockA{
        Departamento de Ciência da Computação\\
        Universidade Federal de Minas Gerais\\
        Belo Horizonte, Brasil\\
        flavioro@dcc.ufmg.br
    }
    \and
    \IEEEauthorblockN{Luiz Chaimowicz}
    \IEEEauthorblockA{
        Departamento de Ciência da Computação\\
        Universidade Federal de Minas Gerais\\
        Belo Horizonte, Brasil\\
        chaimo@dcc.ufmg.br
    }
}

    
    

\maketitle

\begin{abstract}
Pixel art is a popular artistic style adopted in the gaming industry, and nowadays, it is often accompanied by modern rendering techniques. One example is dynamic lighting for the game sprites, for which normal mapping defines how the light interacts with the material represented by each pixel. Although there are different methods to generate normal maps for 3D games, applying them for pixel art may not yield correct results due to the style specificities. Therefore, this work compiles different normal map generation methods and study their applicability for pixel art, reducing the scarcity of existing material on the techniques and contributing to a qualitative analysis of the behavior of these methods in different case studies.

\end{abstract}



\section{Introduction} \label{Introduction}
Within the context of gaming platforms, pixel art was born from the limitations of the existing hardware on the first generations of digital games. Those limitations drove developers to adapt and utilize a compatible artistic direction that used two-dimensional and low-resolution artwork, where each pixel on canvas was intentionally and carefully positioned~\cite{silber2015pixel}.

With the available hardware, three-dimensional rendering technologies were introduced along with new techniques that promoted complex visual effects, like normal mapping for dynamic lighting of surfaces. The normal mapping technique encodes normal vectors on a texture, which are mapped on a geometry shape and finally utilized to define the behavior of light interactions with the surface of an object. Thus, the rendering engine considers the additional information stored in the texture when it generates the dynamic illumination effects.

Traditionally utilized on three-dimensional games, the normal mapping technique was eventually employed in pixel art. This new style then became known as \emph{Hi-bit pixel art}, mixing retro-style graphics with more modern graphical techniques~\cite{hibitPixelart}. Thus, pixel art transitioned from a consequence of technology limitations into an art direction choice nowadays~\cite{whyPixelArtBecomePopular}.


A normal map encodes at each pixel the normal vector of the surface geometry represented in the tangent space~\cite{akenine2019real}. The RGB color components store the $(x,y,z)$ and, as an example, using 8-bit channels, the color $(128, 128, 255)$ \swatch{bluish} represents a normal pointing upwards $(0, 0, 1)$. Because most normal vectors are either exactly or a small perturbation of that direction, normal maps typically have a bluish tone.

\begin{figure}[ht]
\centerline{\includegraphics[width=\allimgwidth]{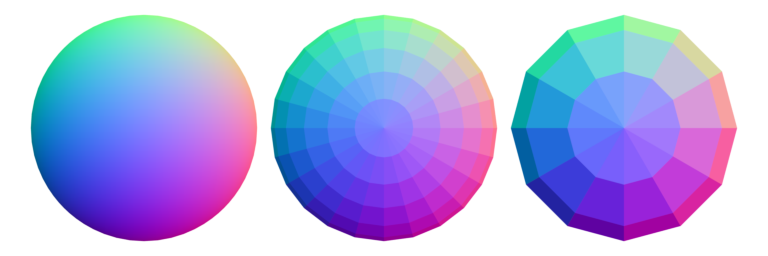}}
\caption{Example of normal map used as reference for manual painting~\cite{ManualPaintAsepriteImageRef}.}
\label{manual_normal_map_ref}
\end{figure}

A good normal map in the context of pixel art encodes the correct geometric information that allows a viewer to infer the objects' depth when rendered using dynamic lighting. Properly validating the quality of a normal map requires the keen eyes of artists in an interactive system that renders the illuminated objects with a moving light source. Alternatively, an approximate form of validation is to compare the colors in the generated map with the ones in reference normal map images, such as in Fig.~\ref{manual_normal_map_ref}.

Given the possibility of exploring three-dimensional techniques in a bi-dimensional context, the existence of various methods to generate normal maps for 3D games, and the absence of compiled material regarding those different techniques \rbtadd{in the literature}, this paper surveys the existing methods \rbtadd{for normal map generation} and assesses their applicability for the specific case of pixel art, analyzing their strengths and weaknesses on common game development use cases.

With the conclusion of our work, we obtained a convenient compilation of material along with explanations about six methods of normal map generation. We conducted experiments to assess the applicability of those methods and summarized the results highlighting each technique's suitability for specific pixel art use cases.


\section{Normal Map Generation} \label{Normal_Map_Generation}
The most common methods for generating a normal map stem from different ideas, ranging from manual painting of these maps to the use of tools that automate their generation. Next, we present six techniques.

\subsection{Hand painting} \label{Manual_painting}
Hand painting is done using reference normal maps as seen in Fig.~\ref{manual_normal_map_ref}, which has examples of geometries projected onto a plane to help the artist visually translate the reference's information into the crafted map. Thus, it depends on the artist's visualization to identify the shape of the entities in the sprite and his experience on materializing this depiction on the canvas using the samples provided by the reference.

A proper visualization of the normal map requires interactive rendering with movement between the light source and the illuminated objects. Due to limitations in the adopted media of this article, we chose to render the results with a single static light source located in the upper-right corner of the images to standardize the shading conditions among the examples.

As a workflow example, it is possible to separate the painting process into steps, starting with vertical faces, followed by horizontal faces, then angled faces, and finally by painting and merging extensive and less detailed groups of normals on the surfaces close to the edges~\cite{Add2DDepthManualPainting}.

\subsection{Sobel filter from a color map} \label{Sobel_filter_from_a_color_map}
The method using a Sobel filter over a color map consists of converting the colored image to grayscale, followed by convolutions with the difference matrices from the Sobel filter, once per axis (\emph{x} and \emph{y}). The two resulting images store a single-channel equivalent to the partial derivative on each axis. A normal map is generated by merging horizontal and vertical derivatives in the first two channels (red and green) while using the third (blue) as a parameter that controls the intensities of gradients. Such a parameter allows controlling the angles of the normal vectors encoded in the generated map~\cite{NormalMapFromHeightMap}.

\begin{figure}[ht]
\centerline{\includegraphics[width=\threesquareimgwidth]{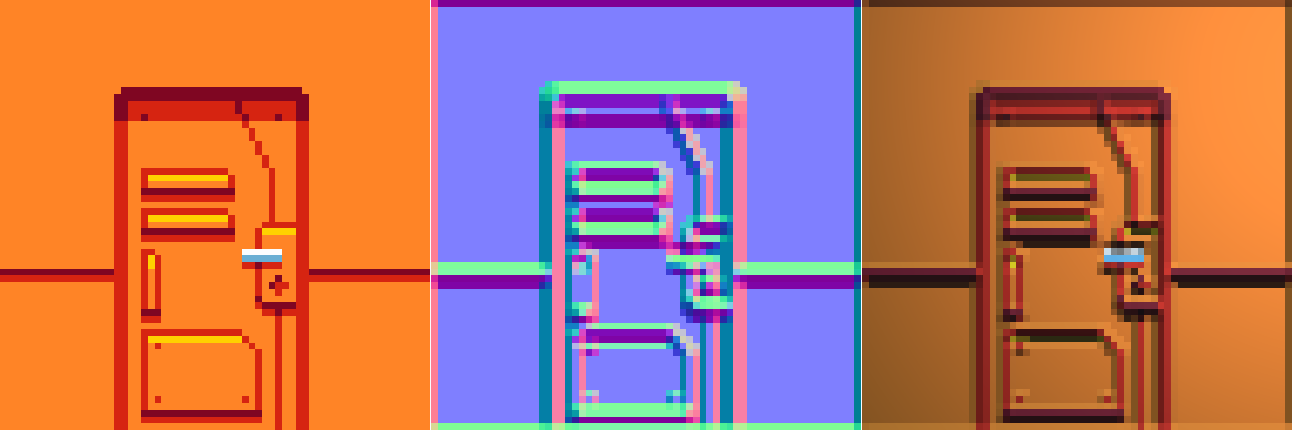}}
\caption{Example of a color map\protect\rbtadd{~(left)} used as a grayscale input for a Sobel filter with its corresponding normal map\protect\rbtadd{~(center)} and rendered image\protect\rbtadd{~(right)}.}
\label{grayscale_albedo}
\end{figure}

In this case, the process for generating the map is less laborious, requiring no intervention by the artist beyond the initial configuration of the intensity parameter for the final map. However, as can be seen in Fig.~\ref{grayscale_albedo}, incorrect and unexpected geometry can be generated from the lighting and shading details that are already present in the image\cite{KatsbitsNoNormalMapFromImage}.

\subsection{Sobel filter from manually painted height map} \label{Sobel_filter_from_manually_painted_height_map}
\begin{figure}[ht]
\centerline{\includegraphics[width=\threesquareimgwidth]{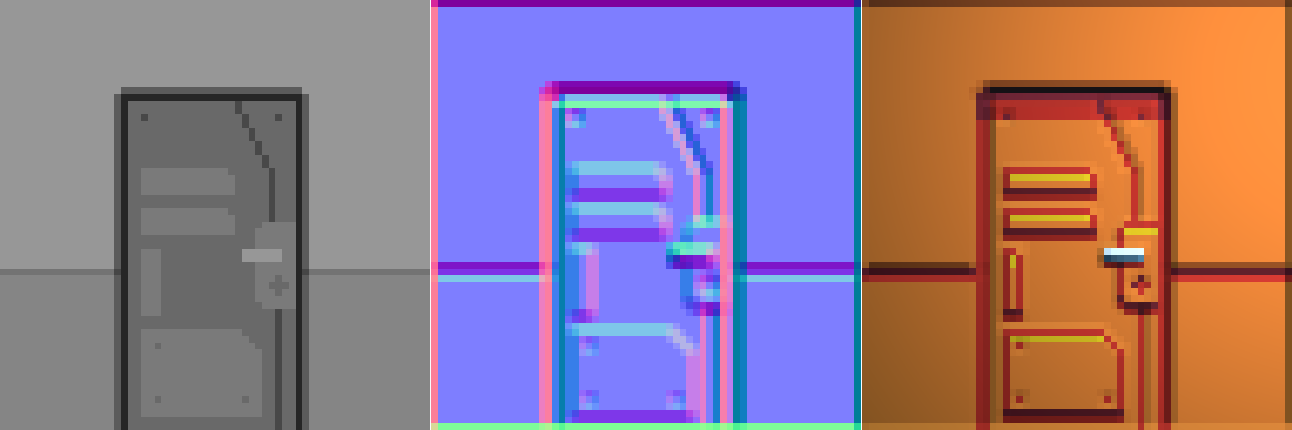}}
\caption{Example of a manually painted height map\protect\rbtadd{~(left)} used as an input for a Sobel filter with its corresponding normal map\protect\rbtadd{~(center)} and rendered image\protect\rbtadd{~(right)}.}
\label{heightmap_albedo}
\end{figure}

This method is similar to the one presented in the previous section. However, instead of using the colored image as input, this method involves creating a height map to be fed to the Sobel operation\rbtadd{~\cite{KatsbitsNoNormalMapFromImage}}. The main difference can be seen in the result after the execution of the technique, as shown in Fig.~\ref{heightmap_albedo}, in which the final render does not contain lighting artifacts as the height map does not have artist-baked shading. However, this technique requires the manual creation of a height map, which can become laborious especially if the number of images is large.

\subsection{Beveling} \label{Beveling}
The beveling method is based on the idea of inflating the surface of the source image based on internal and external contours, as seen in software like Laigter~\cite{laigter} and Sprite Illuminator~\cite{spriteIlluminator}. However, due to the scarcity of literature describing this technique, we propose our implementation from what can be observed in the mentioned software.

\begin{figure}[ht]
\centerline{\includegraphics[width=85mm]{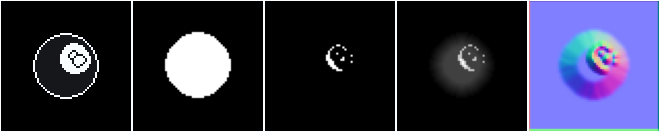}}
\caption{Source image\protect\rbtadd{~(first)}, two binary masks (external and internal shapes) extracted from the source image\protect\rbtadd{~(second and third)}, the distance transform of the two masks multiplied \protect\rbtadd{(fourth)} and the normal map generated from the Sobel filter output\protect\rbtadd{~(fifth)}.}
\label{bevel_normal}
\end{figure}

The proposed algorithm generates a height map from the colored sprite to be used as an input for a Sobel filter similar to the previous techniques. However, this height map is generated automatically by merging the distance transforms~\cite{BevelFromEDT} of two binary masks. These are a transparency mask of the object silhouette and an edge detection mask that encodes internal information about the shapes present in the object. Both are passed through an euclidean distance transform (EDT)~\cite{BevelFromEDT}, which returns an image where each pixel value represents the distance to the nearest edge. They are then merged through a weighted interpolation. Finally, the output of the interpolation is passed through a Gaussian filter that smooths the resulting height map. \rbtadd{This process is exemplified in Fig.~\ref{bevel_normal}.}

The proposed method exposes parameters that allows the fine tuning of the algorithm for each input. It allows the usage of thresholds for clipping the edge detection used to generate the internal shapes binary mask. It also exposes two parameters that intensify the output of the distance transforms, allowing to change the strength of the external and internal generated height maps. Finally, it exposes a parameter that specifies the weight of the interpolation between the EDT outputs for the internal shapes and the external shape.

\subsection{Merge of four illumination angles} \label{Merge_of_four_illumination_angles}
\begin{figure}[ht]
\centerline{\includegraphics[width=77mm]{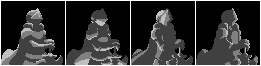}}
\caption{Example of four illumination angles from a knight sprite (\protect\rbtadd{from left to right: }top, bottom, left and right light directions).}
\label{four_ilum_knights}
\end{figure}

Merging artist-shaded versions of four lighting angles is the main technique used in Sprite Lamp~\cite{spritelamp}, where a normal map is generated from four images such as in Fig.~\ref{four_ilum_knights}. As it is a commercial and closed-source tool, there is no public material referring to the technique used, but a similar technique is described by Ryan Clark~\cite{RyanC4IlumNormalMap}. He addresses a method that uses four grayscale images with objects illuminated from the upper, lower, left, and right sides. Then, images whose light sources are in different axis (up-left and lower-right) are grouped in different color channels with different color ranges (0 to 127 for the first and 128 to 255 for the latter). Finally, the two resulting images are merged through an overlay operation and comprise the red and green components. The blue color channel is filled according to the user's needs, with higher values resulting in a smoother normal map.

Compared to other partially automated techniques, this method is more laborious for the artist since, for each colored sprite, four others must be drawn with different lighting conditions to generate a normal map. But on the other hand, it offers the artist reasonable control over the final results using concepts that artists are likely familiar with: shading the depicted objects according to some light direction.



\subsection{Generated from deep generative models} \label{Neural_Network_model}
\begin{figure}[ht]
\centerline{\includegraphics[width=\threesquareimgwidth]{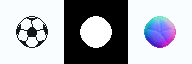}}
\caption{Example of a soccer ball line art\protect\rbtadd{~(left)}, alpha mask\protect\rbtadd{~(center)} and the deep generative model~\cite{DeepNormalEstimation} output\protect\rbtadd{~(right)}.}
\label{soccerball_deepnormals}
\end{figure}

In \cite{DeepNormalEstimation}, the authors propose a deep generative model that receives two images as input, one containing the contours that define the shapes of the object of interest on the sprite and one containing the object's transparency mask. The model then outputs a normal map of the source image, with a proposed geometry similar to the beveling method. So, for images in which the object already has the contour explicitly drawn, this method becomes quite productive, since the extraction of these contours can be simpler than meeting the conditions required by some of the previous methods. However, the used model was trained with a higher resolution, non-pixel art dataset and does not have configuration parameters to control normal intensity. \rbtadd{An example of its application in a pixel art image can be seen on Fig.~\ref{soccerball_deepnormals}.}


\section{Experiments} \label{Experiments}
When generating normal maps for 3D games, the object's geometry and texture are separated; hence, the normal maps do not need to encode macrogeometry information~\cite{akenine2019real} (surface relief represented by vertices and triangles). On the other hand, in pixel art, the object geometry has to be represented by the normal map. In addition, both volume and texture are encoded together in pixel art. To illustrate the challenges, Fig.~\ref{bevel_normal} shows a pool ball made of a very smooth spherical surface with a texture of the ball number. In this case, a correct normal map should consider only the geometry, not the texture. But in the situation of a soccer ball (Fig.~\ref{soccerball_deepnormals}) the normal map must also account for the object's texture, for the seams between patches to be adequately lit.

Adding to the problem, as pixel art images do not have a separate axis for depth information, artists resort to shading and perspective to depict it. In particular, pre-baked shading typically crafted by artists can impact the resulting normal map to some degree, depending on the technique used.

\begin{figure}[ht]
\centerline{\includegraphics[width=\allimgwidth]{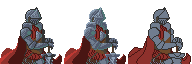}}
\caption{Test cases used on the experiments. From left to right, the outlined, the non-outlined, and the non-shaded sprite.}
\label{test_cases}
\end{figure}

We adopted three test cases \rbtadd{(Fig.~\ref{test_cases}) }to evaluate the performance of the techniques that directly take the color map as its source image, such as the Sobel filter from color map and the beveling method. Those techniques are qualitatively evaluated on three test cases: outlined, non-outlined, and non-shaded sprite. The different outline conditions enable the identification of how well each technique can interpret the inner and outer edges of the object of interest while presenting a satisfactory approximate geometry. As for the absence of lighting and shading, it helps evaluate how sensitive each technique is to these conditions, and their behavior for each case.

For the other techniques, we adopted the outlined sprite since they do not take as input the color map directly. In addition, we used the map in Fig.~\ref{res_heightmap} as input to the Sobel filter from a manually painted height map, while for the different angles technique, we used the four images shown in Fig.~\ref{four_ilum_knights}. Finally, to execute the deep generative model, we manually painted the object's contour and utilized it as shown in Fig.~\ref{res_deepnormals}.

As for the tools used in the experiments, we manually painted the normal maps in the Aseprite image editor\cite{aseprite}. We coded the algorithms in Python for the techniques with Sobel (from the color and the height maps) and the beveling method. As for the technique with different lighting angles, we created the four shaded sprites in Aseprite and fed them to Sprite Lamp~\cite{spritelamp}. For the last technique, we used the pre-trained model from the DeepNormals repository~\cite{DeepNormalEstimation} and provided a hand-crafted edge-only image and a shape mask to the network. Finally, we also created an interactive renderer to visualize the dynamically illuminated images through the Godot game engine~\cite{godot}. The source code and related files have been made available\footnote{
Repository: https://github.com/rodrigodmoreira/nmap-generation}. 

\section{Results} \label{Results}

Next, we present the results for each analyzed technique.

\subsection{Hand painting}
\begin{figure}[ht]
\centerline{\includegraphics[width=\allimgwidth]{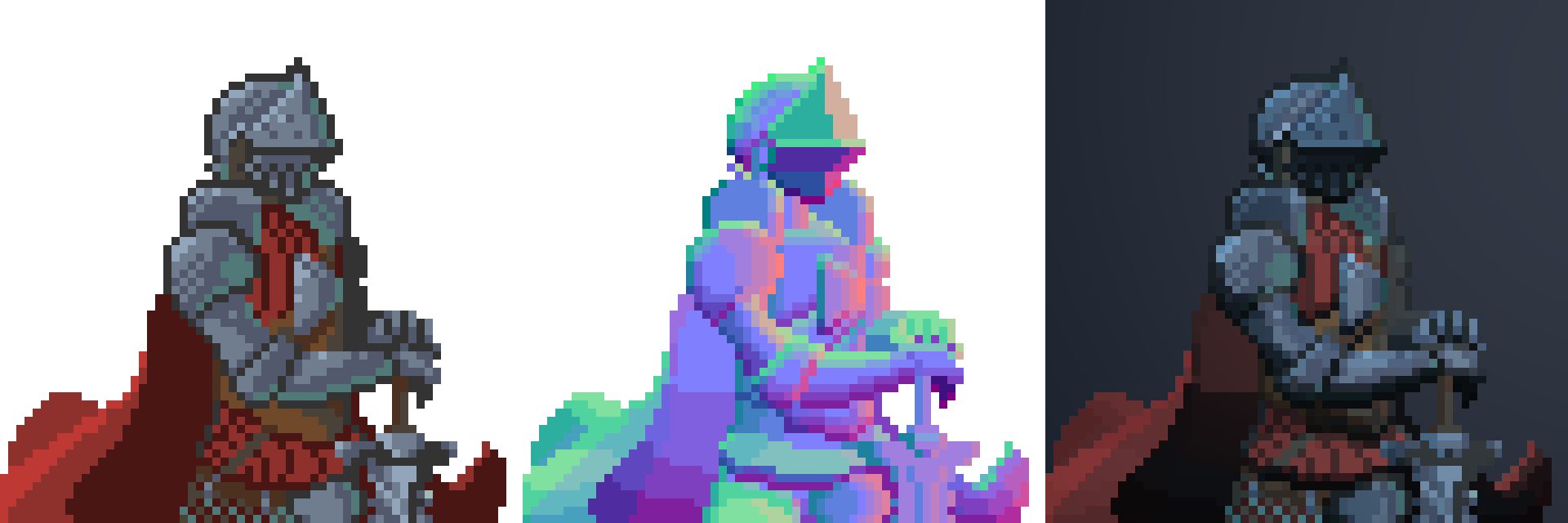}}
\caption{Original sprite\protect\rbtadd{~(left)}, hand painted normal map\protect\rbtadd{~(center)} and the final rendered image\protect\rbtadd{~(right)}.}
\label{res_manual}
\end{figure}

The normal map crafted using the hand painting technique can be seen in Fig.~\ref{res_manual}. As the artist is responsible for manually determining the normal at each pixel, this technique allows full control in the generation. To illustrate, the artist could separate the helmet in a frontal and rear segments, even when the colored sprite had only a very subtle inner edge between them. As this method generates a map closer to the user's intentions, the results obtained here are used as a reference for comparison with the other techniques.

\subsection{Sobel filter from a color map}
\begin{figure}[tp]
\centerline{\includegraphics[width=\allimgwidth]{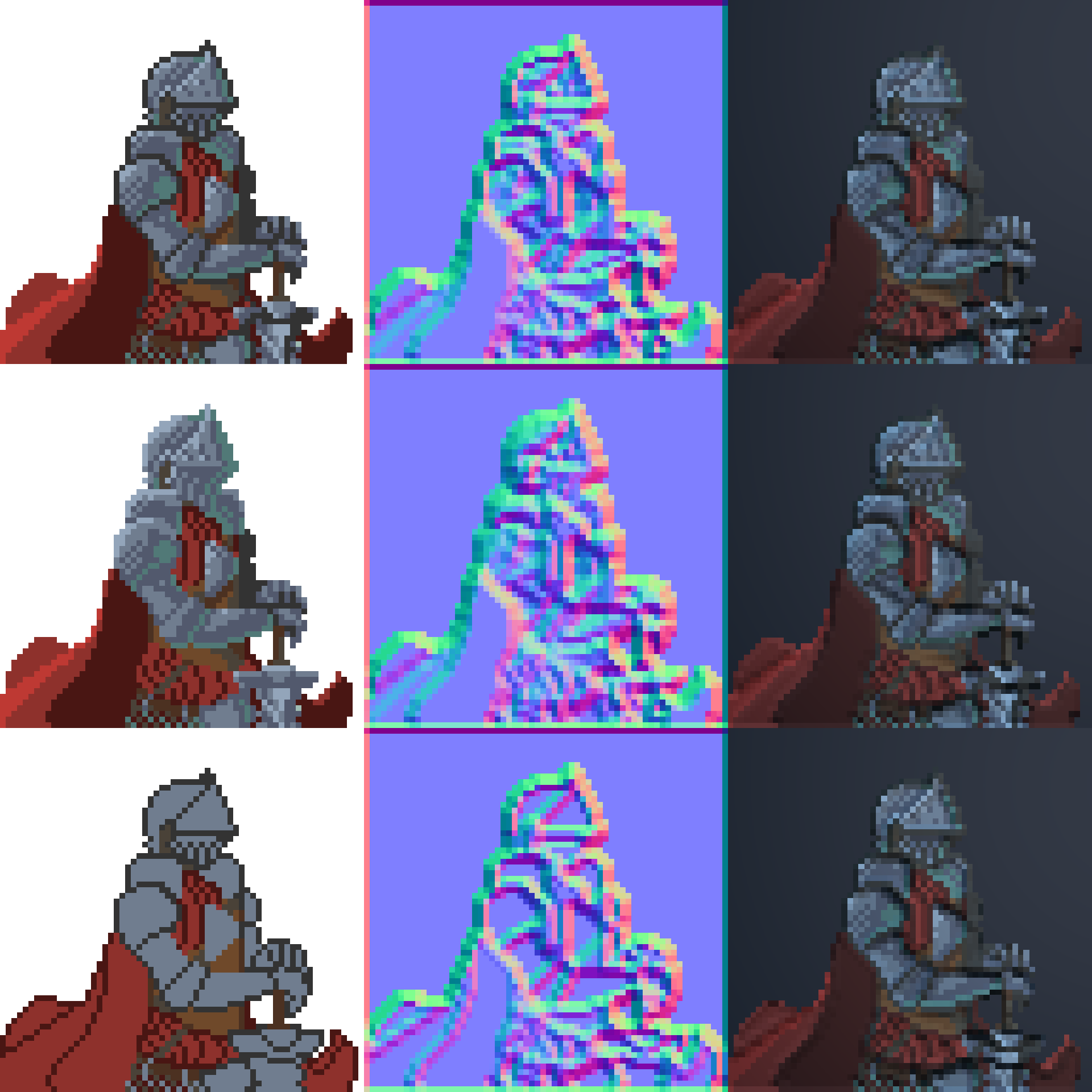}}
\caption{Normal maps\protect\rbtadd{~(center)} generated using Sobel filter from their color map\protect\rbtadd{~(left)}, and the final rendered image\protect\rbtadd{~(right)}. Shaded color map on first row, non-outlined color map on second row and non-shaded color map on third row.}
\label{res_albedo}
\end{figure}

In the technique with Sobel filter from a color map, the obtained normal maps are represented in Fig.~\ref{res_albedo}. For the outlined image (first row), we note that some details are absent, as is the case of the lower and upper parts of the knight's cloak with the final geometry looking as if it's a single face pointing towards the same direction. Another incoherent result can be seen in the right arm, where the innermost part appears as a groove on the surface due to shading, making it seem that it has an inverted volume. One can also notice the lack of more refined control of the geometry, as can be seen on the front of the helmet.

For the non-outlined input (second row), the final result does not have such a great emphasis on some edges, as can be seen near the helmet and the shoulder pad. It also shows the same anomalies seen in the first version of the image, with the inverted volume on the inside.

As for the non-shaded input (third row), a clean normal map is obtained with a more readable result, becoming easier for the observer to define what and where are the internal shapes. However, it has the same anomalies as with the other inputs, with some normals indicating grooves inside the silhouette.

Compared to the hand-painted map, the maps generated via Sobel applied to a color map only show geometry variations close to the detected edges, with the surface direction defined according to the luminosity from the colors near these edges. This contrasts with the hand-painted map, that has the information better distributed along the visible surfaces.

\begin{figure}[ht]
\centerline{\includegraphics[width=\allimgwidth]{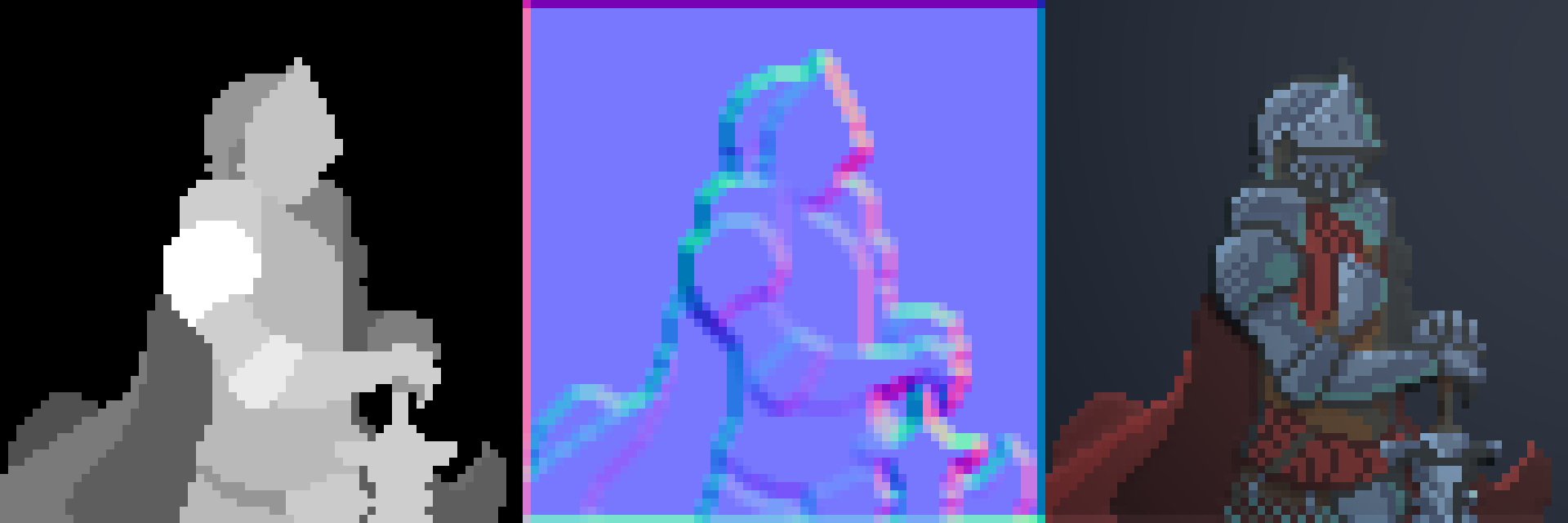}}
\caption{Hand painted height map\protect\rbtadd{~(left)}, it's generated normal map\protect\rbtadd{~(center)}, and the final rendered image\protect\rbtadd{~(right)}.}
\label{res_heightmap}
\end{figure}
\subsection{Sobel filter from a manually painted height map}
The normal map obtained with the Sobel filter on a hand-painted height map can be seen in Fig.~\ref{res_heightmap}. In the generated map, the internal information is very simplified, but it captured the overall shape of the knight. This happens because the height map used is very simplified, prioritizing showing the height difference between larger parts of the object. It is also related to the difficulty in generating this map by hand, since the responsibility for visualizing the depth of each pixel and transporting this visualization to the canvas is transferred to the artist. 

Compared to the hand-painted normal map, most of the geometry details are found near the edges, without a good distribution of information along the visible surfaces of the object. However, unlike the Sobel applied directly to the color map, here, no grooves are present since the original image contains explicit information about the depth of the objects in the image without the interference of lighting and shading conditions that were already encoded in the original sprite.

\subsection{Beveling}
\begin{figure}[ht]
\centerline{\includegraphics[width=\allimgwidth]{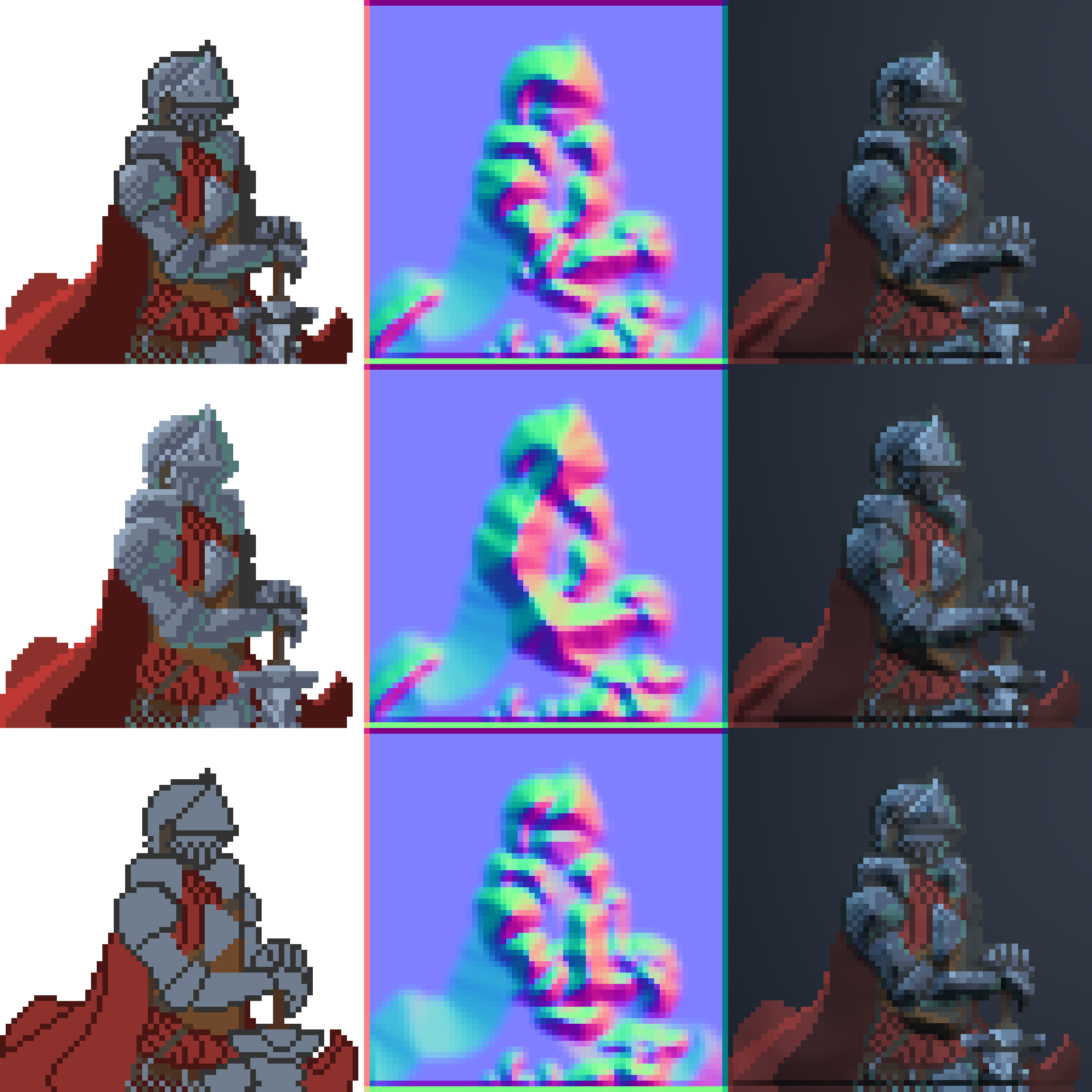}}
\caption{Normal maps\protect\rbtadd{~(center)} generated using beveling from their color map\protect\rbtadd{~(left)}, and the final rendered image\protect\rbtadd{~(right)}. Shaded color map on first row, non-outlined color map on second row and non-shaded color map on third row.}
\label{res_bevel}
\end{figure}

The normal maps from the beveling technique are represented in Fig.~\ref{res_bevel}. In the outlined image (first row), the generated geometry takes into account both the shape of the silhouette and the information of its internal parts, with no portions of the image containing an inverted volume, as observed on the Sobel from a color map. However, in cases such as the helmet and the lower part of the cloak, we observe diminished control over the details, with the generated geometry different than the expected output.

For the non-outlined image (second row), we notice a reduction of the details inside the silhouette of the knight. Without the inner edges, the algorithm struggles to segment the different parts in the image. The output presents a lower level of internal detail while still maintaining a general influence of the outer shape of the image. It also merges small shapes, grouping them into bigger parts and reducing the noise levels on the normal map. 

For the non-shaded sprite (third row), there's a noticeable noise decrease in the output. It depicts a more generalized geometry that emphasizes the outer shapes of the objects present in the image. However, removing the shading made it difficult for the algorithm to correctly define the arm since the region near the torso has a brown leather piece with its color luminosity sufficiently close to the arm's color, hampering the distinction between the two. 

Compared to the result via manual painting, the geometry effect generated by the map via bevel presents a geometry much less faithful to the original image, as can be seen on the helmet. In addition, the generated height map has portions of the image identified as separate objects that do not correspond directly to an existing element in the original sprite, which can be seen on the cloak for the shaded inputs. Finally, parts that require a geometric effect that simulates a negative volume towards the surface are not present in the final map and can only be seen in the hand-painted technique.

\subsection{Merge of four illumination angles}
\begin{figure}[ht]
\centerline{\includegraphics[width=\allimgwidth]{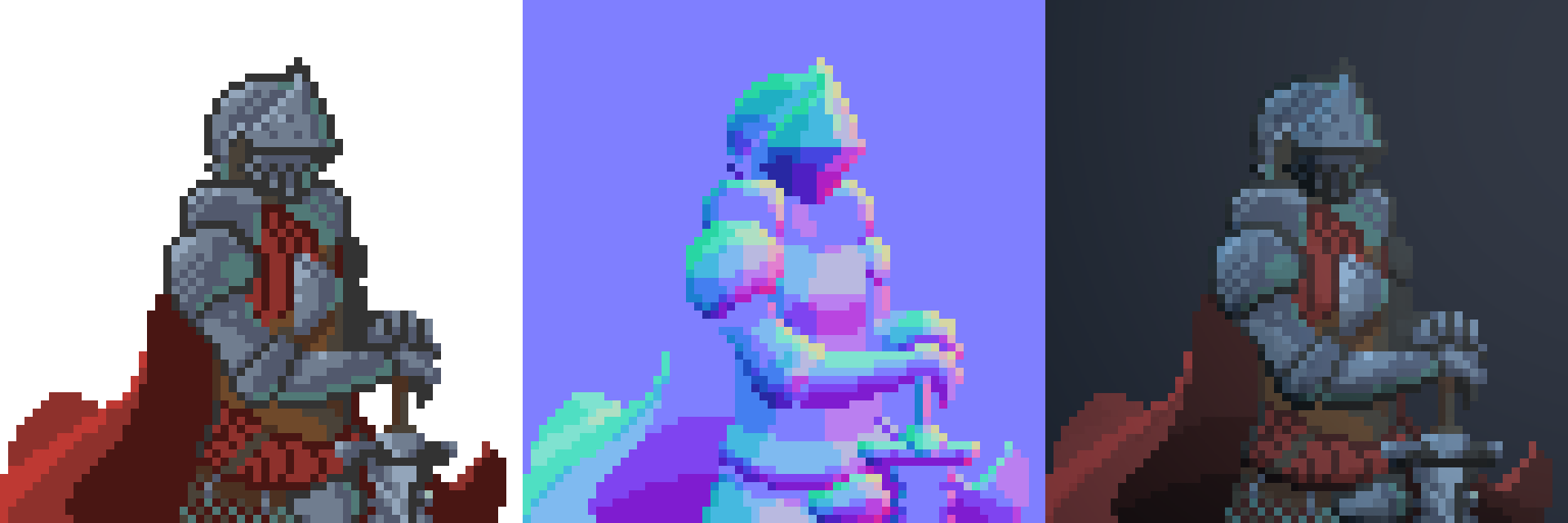}}
\caption{Original sprite\protect\rbtadd{~(left)}, normal map\protect\rbtadd{~(center)} generated from manually painted sprites with 4 different angles (Fig.~\ref{four_ilum_knights}), and the final rendered image\protect\rbtadd{~(right)}.}
\label{res_4ilum}
\end{figure}

Fig.~\ref{res_4ilum} shows the test case with the technique that requires four illuminated sprites in different angles. This method allowed great control over the normals with a smoother distribution along the surfaces, as seen on the lower surface of the cloak. The generated map also presents the expected geometry as can be seen on the helmet.

Compared with hand-painted normal maps, the generated maps were fairly similar, with the hand-painted output presenting a higher level of detail.

\subsection{Generated from a deep generative model}
\begin{figure}[ht]
\centerline{\includegraphics[width=\allimgwidth]{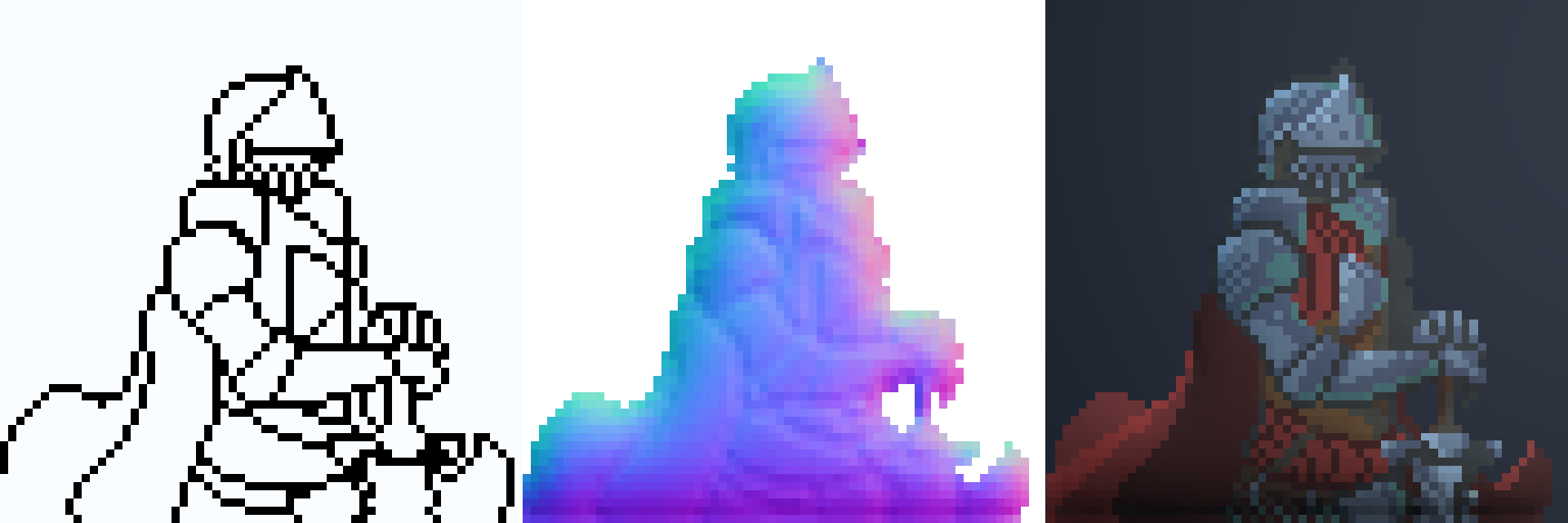}}
\caption{Outline of the original image\protect\rbtadd{~(left)}, its normal map generated from the deep generative model estimation\protect\rbtadd{~(center)}, and the final rendered image\protect\rbtadd{~(right)}.}
\label{res_deepnormals}
\end{figure}

As for the usage of the deep generative model, the generated normal map can be seen in Fig.~\ref{res_deepnormals}. The geometry is close to the expected but with less intense normals and blurry and imprecise edges. In addition, an exaggerated smoothing of the internal elements is noticeable, making the geometry effect very subtle in the rendered image. This might be a consequence of the training data domain of the model, since it learned with higher resolution images, not pixel art.

Compared to the results from the hand-painted technique, it presents very subtle details in the internal region while emphasizing the external shape, with the latter not being so pronounced in the hand-painted result. It also shows a smoother transition among the generated normals.

\section{Discussion} \label{Discussion}
Based on the results\footnote{Video: https://rodrigodmoreira.github.io/nmap-generation-video/}, we can see that the generated normal map may present different levels of noise and information in its interior depending on the method of choice and the artistic style of the input image while requiring different levels of labor from the artist. Furthermore, none of the automated techniques could deliver a completely satisfactory geometry, as is the case with the knight's helmet. Only the hand-painted normal map and hand-painted four-light angles showed a geometry consistent with the expected, at the expense of greater amounts of work and extra complexity involving the creation of the required inputs.

For the techniques dependent on a color map as input, the absence of well-defined contours could generate a less precise edge detection on each method results, leading to a consequent reduction of information within the silhouettes. Furthermore, the presence of lighting and shading details previously drawn in the input image introduces noise in the final map, which may hamper the identification of smaller details in the final map. Ideally, inputs with contrasting edges and without lighting and shading can provide a final result with greater clarity in existing details and better precision when generating the expected geometry for smaller objects.

Lastly, based on the research and results obtained, we distinguish the surveyed techniques as follows:
\begin{itemize}
    \item \textbf{Sobel from a color map} uses information exclusively from edges, resulting in incoherent geometry. While it doesn't yield good results on general pixel art, it might still produce acceptable results in texture-like sprites (e.g., grass, water, wood) where noise can be perceived as an intentional resource;
    \item \textbf{Sobel from a height map} requires the hand-painting of a height map, but it allows the modeling of the geometry and texture of the surface depending on the level of detail from the input. It yields good results on pixel art but may present a higher difficulty for the artist to paint the depth of each pixel. It can also prevent artifacts from pre-baked shading;
    \item \textbf{Beveling} doesn't require the creation of new images and can separate the internal shapes of the object, achieving better results with internal contours. It produces good results for objects that we expect to have a beveled geometry, being a useful technique for non-planar elements;
    \item \textbf{Four-angles of lighting} requires the hand-painting of four images for each sprite and can model the geometry and (optionally) the texture of the surface. It requires more labour than the Sobel from height map method, but it uses concepts that artists are better acquainted with, such as drawing objects under different lighting conditions. It's a good option for sprites in general, but the quality depends on the angled illuminated images drawn;
    \item \textbf{Deep generative model} is fully automated but it might produce weak internal geometry on pixel art. It places weak creases on the edges and doesn't separate well parts of the object. In pixel art, it may be more useful on objects mainly composed of a single and simple outer shape and that doesn't require much internal information, like a ball. 
\end{itemize}

\section{Conclusion} \label{Conclusion}
This work surveyed six techniques for generating a normal map and analyzed its applicability in the context of pixel art. We found that no technique could produce maps as good as the one an artist can manually paint, but depending on the characteristics of the sprites, different methods might get closer to the expected result. Compiling these techniques is an initial contribution toward the reduction of the scarcity of available material, while easing the access to information regarding their inner workings. It was also possible to identify the sensitivity of some of these techniques to the given inputs and compare the strengths and weaknesses of each one over the analyzed test cases. Lastly, through experimentation, it was possible to propose an algorithm for the beveling technique since it did not have any documentation of its execution.

For future work, there is still space to explore new techniques not addressed here, in addition to expanding the number of case studies for each method, giving a better understanding of the behavior of each technique under different artistic styles. Finally, another subject of study is the training of a deep generative model on a pixel art dataset. 

\bibliographystyle{IEEEtran}
\bibliography{IEEEabrv,main}

\end{document}